\documentclass[11pt]{article}
\usepackage{geometry}
\usepackage{graphicx}
\usepackage{amssymb}
\usepackage{epstopdf}
\usepackage{latexsym}
\usepackage{subfigure}
\usepackage{float}
\begin{document}

\begin{titlepage}
\begin{flushright}
FR-PHENO-2010-031
\end{flushright}

\bigskip
\bigskip
\bigskip

\begin{center}

{\Large \bf Self-Compactifying Gravity}

\vspace{0.5cm}

{\bf Beyhan Puli\c{c}e$^1$ and \c{S}\"ukr\"u Hanif Tany\i ld\i z\i$^2$}

\vspace{.8cm}

$^1${\it Institut f\"ur Physik, Albert-Ludwigs Universit\"at Freiburg
Hermann-Herder-Str. 3, 79104 Freiburg i.B., Germany}

$^2${\it Bogoliubov Laboratory of Theoretical Physics, Joint Institute for Nuclear Research, 141980, Dubna, Moscow Region, Russia}

\end{center}
\vspace{1cm}
\begin{abstract}
\medskip

We study the self-compactification of extra dimensions via higher curvature gravity, $f(\mathcal{R})$, where
$f(\mathcal{R})$ is the generic function of the Ricci scalar $\mathcal{R}$. First, we reduce pure $f(\mathcal{R})$ theory to a scalar-tensor theory by a conformal
transformation \cite{fr-einstein, barrow}. Then we show that, by a second conformal
transformation, this scalar-tensor theory turns out to a
nonminimal scalar-tensor theory. We find non-vanishing scalar field configurations that
satisfy the conditions on the partially vanishing energy-momentum tensor and the equations of motion of the nonminimal scalar-tensor theory. It is interesting that we find that the source of gravity, $\phi$, has discrete spectrum.
The minimum of the potential changes according to the value of the coupling constant of the scalar field to the curvature scalar.
When the minimum is at zero for a vanishing scalar field, the entire spacetime is flat. When the minimum is at a nonzero value for a non-vanishing scalar
field, the extra space is compactified. We thus show that a given $f(\mathcal{R})$ theory can self-compactify the extra dimensions.

\end{abstract}

\medskip


\end{titlepage}

\section{Introduction}
\label{intro}

The theories which contain only generic $f(\mathcal{R})$ function
in the lagrangian is mathematically equivalent to Einstein gravity plus a scalar field theory
\cite{fr-einstein, maeda, barrow}.

The scalar-tensor theories, \emph{i.e.} Einstein gravity plus a scalar field theories, 
may have scalar fields
which couple to gravity minimally or nonminimally. Dicke
discussed \cite{dicke} the conformal transformation from
Brans-Dicke theory \cite{brans-dicke} to the minimally coupled
case. Conformal transformations relate a nonminimal
scalar-tensor theory to a minimal scalar-tensor theory
\cite{bekenstein}.

In order to reduce $f(\mathcal{R})$ theory to a scalar-tensor
theory with the coupling of the scalar field with the curvature scalar (or Ricci scalar; 
we will use both of these terms), we use conformal transformations; in other words we rescale
the metric tensor. The rescaled metric tensor is not a redefinition of the
metric tensor in a different coordinate system; each of the metric tensors
describes different gravitational fields and different physics.

Spontaneous compactification has been discussed using higher order invariant terms of the curvature tensor
\cite{wetterich}, Einstein gravity with the coupling to matter \cite{cremmer}, antisymetric tensor fields
\cite{freund} and Yang-Mills fields
\cite{luciani, randjbar}. Spontaneous compactification mechanisms has also been
analyzed with sigma models \cite{gell-mann} and
conformally-coupled scalars \cite{gerard}. It has been shown that for some toy models, dynamical 
compactification is realized if all supersymmetries are spontaneously broken \cite{dvali-shifman}.
Spontaneous compactification of Lovelock theory in vacuum has been
shown in \cite{canfora}.

Here, we give a brief summary of a spontaneous compactification
mechanism via a single scalar field \cite{demir} to make a
complete discussion in our work. In \cite{demir}, they consider a
scalar-tensor theory which has a nonminimal coupling term that
shows the coupling of the scalar field with the Ricci scalar. They
assume some specific conditions for the source term of the Ricci
tensor so the scalar field gravitates only in a subset of
dimensions. They find the scalar field configurations and the
corresponding self-interaction potential that satisfy these
conditions and the equations of motion of the theory (which was
also found in \cite{ayon-beato}). They find specific scalar field
configurations that do not gravitate in four-dimensional spacetime
but only in extra space. When the vacuum expectation value of the
scalar field is zero the entire space is completely flat but when
it has a nonzero value, the extra dimensions are compactified
while the four dimensions remain flat. The resulting topology
becomes the ordinary four-dimensional spacetime times the compact
manifold of extra dimensions, $M^{4}\otimes E^{d}$.

Our aim is to discuss a self-compactification mechanism via a more
general scalar field in a scalar-tensor theory which is induced by 
$f(\mathcal{R})$ theory. We make two conformal
transformations; applying the first one we obtain Einstein gravity
plus a scalar field theory that does not contain any nonminimal terms
(minimal scalar-tensor theory) from $f(\mathcal{R})$ theory, then by
applying the second one we reach a scalar-tensor theory that
includes various nonminimal terms (nonminimal scalar-tensor theory).

Considering the resulting nonminimal scalar-tensor theory, we study 
self compactification of extra dimensions via a single
scalar field which lives in the entire spacetime however
gravitates only in a subset of dimensions. We find the scalar
field configurations that fulfills the conditions on the energy
momentum tensors.

In Sec. \ref{fr}, we begin our analysis by considering
$f(\mathcal{R})$ theory. We apply a conformal transformation to
transform this theory to a minimal scalar-tensor theory. Then by applying
another conformal transformation, we obtain a nonminimal scalar
tensor theory. This theory has a nonminimal kinetic term and a
nonminimal coupling of the scalar field with the curvature
scalar.

In Sec. \ref{self}, we study the self-compactification
mechanism. First, we find the energy momentum tensor of the
scalar field, $\phi$. In Sec. \ref{subself1}, we want that all
components of the source term of the Ricci tensor vanish in four dimensions in
order to have a scalar field that does not gravitate in
four-dimensional spacetime. In this manner, we find the explicit
form of the scalar field and the corresponding self-interaction potential. In Sec. \ref{subself2}, we analyze the compactification mechanism.
We want the pure extra-dimensional components of the source term of the Ricci
tensor not vanish. Hence, we find a non-vanishing 
curvature scalar of the extra dimensions since the curvature tensor is proportional to its
source term. This non-vanishing curvature scalar means extra dimensions are curved while ordinary four dimensions remain flat, so there is 
a compactification effect. This is self-compactification since we derive the nonminimal scalar-tensor theory from $f(\mathcal{R})$ by conformal
transformations, \emph{i.e.} $f(\mathcal{R})$ itself gives rise to compactification of extra dimensions.

\section{From $f(\mathcal{R})$ Theory to Scalar-Tensor Theory}
\label{fr}

We know that, as a direct generalization of the Einstein-Hilbert
theory, the modified gravity theories by a generic function
of the curvature scalar, $f(\mathcal{R})$, are equivalent to
Einstein gravity (with the same fundamental scale) plus a scalar
field theory. We derive a minimal scalar-tensor theory from $f(\mathcal{R})$ applying a
conformal transformation. We will show that the theory constructed, following
another conformal transformation, turns out to be a model with nonminimal
coupling that realizes the coupling between the curvature scalar and
the scalar field. The kinetic term of this nonminimal scalar-tensor theory has 
some function of $\phi$, $F(\phi)$, as a factor.
First of all, we consider the action of $f(\mathcal{R})$ theory
\begin{eqnarray}
\label{actionfr} S=\int d^Dx \sqrt{-g}M^{D-2}_\star
f({\mathcal{R}})
\end{eqnarray}
where $M_\star$ is the fundamental scale of gravity and $f(\mathcal{R})$ is a generic function of
the curvature scalar $\mathcal{R}$. We remind that the chosen function of $\mathcal{R}$ should be conformal invariant. We find the Einstein tensor by taking variation of (\ref{actionfr}) 
with respect to the metric $g_{AB}$
\begin{eqnarray}
\label{motioneqfr}
G_{AB}&=&\mathcal{R}_{AB}-\frac{1}{2}g_{AB}\mathcal{R}\nonumber\\
&=&\left[f^\prime(\mathcal{R})\right]^{-1}\Bigg\{\frac{1}{2}g_{AB}f(\mathcal{R})-\frac{1}{2}g_{AB}f^\prime(\mathcal{R})\mathcal{R}-g_{AB}\Box
f^\prime(\mathcal{R})+\nabla_A\nabla_Bf^\prime(\mathcal{R})\Bigg\}
\end{eqnarray}
where
\begin{eqnarray}
f^\prime(\mathcal{R})\equiv\frac{\partial
f(\mathcal{R})}{\partial\mathcal{R}}~~~.
\end{eqnarray}
Then, we make the conformal transformation                         
\begin{eqnarray}
\label{trans1} 
\hat{g}_{AB}(z)=e^{2\Phi(z)} g_{AB}(z)~~~.
\end{eqnarray}
Considering this conformal transformation, we find the connection coefficients 
\begin{eqnarray}
\label{connection}
\hat{\Gamma}^C_{AB}=\frac{1}{2}\hat{g}^{CD}\Big(\partial_A\hat{g}_{BD}+\partial_B\hat{g}_{DA}-\partial_D\hat{g}_{AB}\Big)~~~.
\end{eqnarray}
Substituting these new connection coefficients into the following Riemann tensor relation
\begin{eqnarray}
\label{Riemanntensorrel}
\hat{\mathcal{R}}^C_{ADB}=\partial\hat{\Gamma}^C_{AB}-\partial\hat{\Gamma}^C_{AD}+\hat{\Gamma}^C_{DE}\hat{\Gamma}^E_{AB}-\hat{\Gamma}^C_{BE}
\hat{\Gamma}^E_{AD}
\end{eqnarray}
we obtain the transformed Riemann tensor in terms of the untransformed Riemann tensor
\begin{eqnarray}
\label{Riemanntensor}
\hat{\mathcal{R}}^C_{ADB}&=&\mathcal{R}^C_{ADB}+\Bigg[\nabla_D\nabla_A\Phi-(\nabla_A\Phi)(\nabla_D\Phi)+(\nabla_E\Phi)(\nabla_G\Phi)g^{EG}g_{AD}
\Bigg]\delta^C_B\nonumber\\
&-&\Bigg[\nabla_B\nabla_A\Phi-(\nabla_A\Phi)(\nabla_B\Phi)\delta^C_D+(\nabla_E\Phi)(\nabla_G\Phi)g^{EG}g_{AB}\Bigg]\delta^C_D\nonumber\\
&+&\Bigg[\nabla_B\nabla_F\Phi-(\nabla_B\Phi)(\nabla_F\Phi)\Bigg]g^{CD}g_{AD}\nonumber\\
&-&\Bigg[\nabla_D\nabla_F\Phi-(\nabla_F\Phi)(\nabla_D\Phi)\Bigg]g^{CF}g_{AB}~~~.
\end{eqnarray}
Then, contracting the Riemann tensor above with the transformed metric (\ref{trans1}), we
get the transformed Ricci tensor
\begin{eqnarray}
\label{confriccitensor}
\hat{\mathcal{R}}_{AB}&=&\mathcal{R}_{AB}-(D-2)\Bigg\{\nabla_A\nabla_B\Phi-\nabla_A\Phi\nabla_B\Phi+g_{AB}g^{CD}\nabla_C\Phi\nabla_D\Phi\Bigg\}
\nonumber\\
&-&g_{AB}g^{CD}\nabla_C\nabla_D\Phi
\end{eqnarray}
and contracting this Ricci tensor with the transformed metric (\ref{trans1}), we obtain the Ricci scalar
\begin{eqnarray}
\label{confricciscalar} \hat{\mathcal{R}}= e^{-2\Phi}
\Bigg\{\mathcal{R}-2(D-1)g^{AB}\nabla_{A} \nabla_{B} \Phi-
(D-1)(D-2)g^{AB}\nabla_{A}\Phi \nabla_{B} \Phi\Bigg\}~~~.
\end{eqnarray}
Substituting relations (\ref{confriccitensor}) and
(\ref{confricciscalar}) into
\begin{eqnarray}
\label{einsteinstrans}
\hat{G}_{AB}=\hat{\mathcal{R}}_{AB}-\frac{1}{2}\hat{g}_{AB}\hat{\mathcal{R}}
\end{eqnarray}
we obtain the conformally transformed Einstein tensor in terms of
the untransformed Einstein tensor (\ref{motioneqfr})
\begin{eqnarray}
\label{einsteintrans}
\hat{G}_{AB}&=&G_{AB}-(D-2)\Bigg[\nabla_A\nabla_B\Phi-g_{AB}\Box\Phi\Bigg]\nonumber\\
&+&(D-2)\Bigg[\nabla_A\Phi\nabla_B\Phi+\frac{D-3}{2}\Big(\nabla\Phi\Big)^2g_{AB}\Bigg]
\end{eqnarray}
We choose $\Phi$ to be
\begin{eqnarray}
\label{scalar1}
\Phi\equiv\frac{1}{D-2}\ln
f^\prime(\mathcal{R})~~~
\end{eqnarray}
in order to have all terms in (\ref{einsteintrans}) in terms of $f(\mathcal{R})$. In other words, a Weyl transformation is performed such that terms
containing derivatives of the scalar cancel. In the transformed model (\ref{actionmin}), in principle, the algebraic equation of motion can be solved for the
scalar. Plugging back the solution into the action results in $f(\mathcal{R})$
gravity. This gives the scalar as a function of the scalar
curvature (\ref{scalar1}). This connection was known, before \cite{maeda, barrow}.

We substitute the derivatives 
\begin{eqnarray}
\label{differentiation1}
\nabla_A\nabla_B\Phi=\frac{1}{D-2}\Bigg\{\frac{f^\prime(\mathcal{R})\nabla_A\nabla_Bf^\prime(\mathcal{R})-\nabla_Af^\prime(\mathcal{R})
\nabla_Bf^\prime(\mathcal{R})}{\left[f^\prime(\mathcal{R})\right]^2}\Bigg\}
\end{eqnarray}
and
\begin{eqnarray}
\label{differentiation2}
\nabla_A\Phi\nabla_B\Phi=\frac{1}{(D-2)\left[f^\prime(\mathcal{R})\right]^2}\nabla_Af^\prime(\mathcal{R})\nabla_Bf^\prime(\mathcal{R})~~~
\end{eqnarray}
into (\ref{einsteintrans}) and obtain
\begin{eqnarray}
\label{newformeinstein}
\hat{G}_{AB}&=&G_{AB}-\frac{\nabla_A\nabla_Bf^\prime(\mathcal{R})}{f^\prime(\mathcal{R})}-\frac{D-1}{2(D-2)}g_{AB}g^{CD}
\frac{\nabla_Cf^\prime(\mathcal{R})\nabla_Df^\prime(\mathcal{R})}{\left[f^\prime(\mathcal{R})\right]^2}\nonumber\\
&+&\frac{D-1}{D-2}\frac{\nabla_Af^\prime(\mathcal{R})\nabla_Bf^\prime(\mathcal{R})}{\left[f^\prime(\mathcal{R})\right]^2}+g_{AB}g^{CD}
\frac{\nabla_C\nabla_Df^\prime(\mathcal{R})}{f^\prime(\mathcal{R})}~~~.
\end{eqnarray}
Here, we make a new scalar field definition to have the Einstein tensor (\ref{newformeinstein}) only in terms
of the new scalar field
\begin{eqnarray}
\label{varphi}
\varphi\equiv 2\Phi.
\end{eqnarray}
So, (\ref{newformeinstein}) becomes
\begin{eqnarray}
\label{einsteineq1}
\hat{G}_{AB}=\frac{1}{4}(D-1)(D-2)\nabla_A\varphi\nabla_B\varphi-\frac{1}{8}(D-1)(D-2)\hat{g}_{AB}\nabla_C\varphi\nabla^C\varphi-\hat{g}_{AB}V(\varphi)
\end{eqnarray}
where the following expression for the potential, $V(\varphi)$, have been introduced
\begin{eqnarray}
\label{potu}
V(\varphi)\equiv\frac{1}{2}e^{-\frac{D}{2}\varphi}\Bigg(\mathcal{R}e^{\frac{D-2}{2}\varphi}-f(\mathcal{R})\Bigg)~~~.
\end{eqnarray}
The kinetic term in (\ref{einsteineq1}) becomes canonical with
a definition of a new scalar field $\phi$ \cite{barrow}
\begin{eqnarray}
\label{phi} \phi\equiv
M_{\star}^{\frac{D-2}{2}}\frac{1}{2}\sqrt{(D-1)(D-2)}\varphi~~~\Longrightarrow~~~\phi=M_{\star}^{\frac{D-2}{2}}
\sqrt{\frac{D-1}{D-2}}\ln f^\prime(\mathcal{R})~~~.
\end{eqnarray}
Hence, the substitution of the differentiation of $\varphi$
\begin{eqnarray}
\label{difphi}
\nabla_A\phi\nabla_B\phi=M_\star^{D-2}\frac{1}{4}(D-1)(D-2)\nabla_A\varphi\nabla_B\varphi
\end{eqnarray}
into (\ref{einsteineq1}) gives the transformed Einstein tensor in terms of the scalar field $\phi$
\begin{eqnarray}
\label{hateinsten}
\hat{G}_{AB}=(\nabla_A\phi)(\nabla_B\phi)-\frac{1}{2}\hat{g}_{AB}(\hat{\nabla}\phi)^2-\hat{g}_{AB}\hat{V}(\phi)~~~.
\end{eqnarray}
We know that the Einstein tensor (\ref{hateinsten}) corresponds to
Eintein gravity plus a scalar field theory \cite{fr-einstein}
\begin{eqnarray}
\label{actionmin} S=\int
d^Dx\sqrt{-\hat{g}}\left\{\frac{1}{2}M_\star^{D-2}\hat{\mathcal{R}}-\frac{1}{2}\Big(\hat{\nabla}\phi\Big)^2-\hat{V}(\phi)\right\}~~~.
\end{eqnarray}
This is a minimally-coupled scalar-tensor theory. $\phi$ couples to gravity via only its kinetic term. We now want to pass a nonminimally coupled scalar-tensor theory to be able to realize vanishing energy-momentum tensor as in \cite{demir, ayon-beato}.
Considering the Einstein gravity plus a scalar field theory (\ref{actionmin}) which is derived from the
higher curvature gravity, $f(\mathcal{R})$, we make the conformal
transformation
\begin{eqnarray}
\label{confmetric} \tilde{g}_{AB}(z)=e^{2\omega(z)}\hat{g}_{AB}(z)
\end{eqnarray}
where 
\begin{eqnarray}
\label{conftrans} e^{\omega(z)}\equiv\Big(1-\zeta
M_\star^{2-D}\phi^2(z)\Big)^{1/(2-D)}~~~.
\end{eqnarray}

The Ricci scalar after the conformal transformation (\ref{confmetric}) is 
\begin{eqnarray}
\label{scalarricci} \tilde{\mathcal{R}}=\Big(1-\zeta
M_\star^{2-D}\phi^2\Big)^{-2/(2-D)}\left\{\hat{\mathcal{R}}-12\frac{D-1}{D-2}\zeta^2
M_\star^{2(2-D)}\phi^2
\frac{\Big(\hat{\nabla}\phi\Big)^2}{\Big(1-\zeta
M_\star^{2-D}\phi^2\Big)^2}\right\}~~~.
\end{eqnarray}
Besides, we need to substitute the following expressions 
\begin{eqnarray}
\label{detmetric} \sqrt{-\tilde{g}}=\Big(1-\zeta
M_\star^{2-D}\phi^2\Big)^{D/(2-D)}\sqrt{-\hat{g}}
\end{eqnarray}
and
\begin{eqnarray}
\label{derivativesquare}
(\tilde{\nabla} \phi)^{2}=(1-\zeta M_{\star}^{2-D} \phi^{2})^{-2/(2-D)}(\hat{\nabla}\phi)^{2}~~~.
\end{eqnarray}
into (\ref{actionmin}).
Finally, after the conformal tranformation (\ref{confmetric}), the theory we have is
Einstein gravity plus a scalar-tensor theory with a nonminimal coupling of the scalar field $\phi$ with
the Ricci scalar $\mathcal{R}$
\begin{eqnarray}
\label{actionnonmin} S=\int
d^Dx\sqrt{-\tilde{g}}\left\{\frac{1}{2}M_{\star}^{D-2}\tilde{\mathcal{R}}-F(\phi)\frac{1}{2}\Big(\tilde{\nabla}\phi\Big)^2-\frac{1}{2}\zeta
\tilde{\mathcal{R}}\phi^2-\tilde{V}(\phi)\right\}
\end{eqnarray}
where
\begin{eqnarray}
\label{confpot} \tilde{V}(\phi)=\Big(1-\zeta
M_\star^{2-D}\phi^2\Big)^{-D/(2-D)}\hat{V}(\phi)
\end{eqnarray}
and
\begin{eqnarray}
\label{funcF} F(\phi)=\frac{1}{1-\zeta
M_\star^{2-D}\phi^2}\left\{-12\frac{D-1}{D-2}\zeta^2M_\star^{2-D}\phi^2+\Big(1-\zeta
M_\star^{2-D}\phi^2\Big)^2\right\}
\end{eqnarray}
is the explicit form of the function in the nonminimal kinetic term. The nonminimal scalar-tensor theory 
(\ref{actionnonmin}) is conformal invariant since it is reduced from a conformal invariant
theory, $f(\mathcal{R})$, by conformal transformations. One should notice that there is a function $F(\phi)$ (\ref{funcF}) which makes the kinetic term nonlinear. We will keep this function and study the compactification mechanism with this form of the action. Here, $\zeta$ is the coupling between the scalar field and the curvature scalar.

\section{Self-Compactification}
\label{self}

We consider a real scalar field $\phi$ living in $D$-dimensional
spacetime with coordinates $z^A=(x^\mu,y^{\overline{i}})$ where
$A=0,1,2,3,\cdots,N,~~ \mu=0,1,2,3,~~i=1,2,3,~~\bar{i}=4,
\cdots,N$. The metric is
$\tilde{g}_{AB}=\eta_{\mu\nu}+\tilde{g}_{{\overline{i}}{\overline{j}}}$,
the metric signature
$\eta_{AB}=\textrm{diag}(-1,+1,+1,+1,\cdots)$, number of
dimensions$=N+1~~\textrm{and}~~D\equiv4+d$.

In this section, we will analyze the compactification mechanism with the action
\begin{eqnarray}
\label{actionnonmin2} S=\int
d^Dx\sqrt{-\tilde{g}}\left\{\frac{1}{2}M_\star^{D-2}\tilde{\mathcal{R}}-F(\phi)\frac{1}{2}\Big(\tilde{\nabla}\phi\Big)^2-\frac{1}{2}\zeta
\tilde{\mathcal{R}}\phi^2-\tilde{V}(\phi)\right\}~~~
\end{eqnarray}
where $M_\star$ is the fundamental scale of gravity and
$\tilde{\mathcal{R}}$ is the curvature scalar. The field configurations
that extremize the action (\ref{actionnonmin2}) satisfy the following equations
of motion
\begin{eqnarray}
\label{motioneq1}
\tilde{\mathcal{R}}_{AB}=\frac{\tilde{\mathcal{T}}_{AB}}{M_{\star}^{D-2}-\zeta
\phi^{2}}
\end{eqnarray}
\begin{eqnarray}
\label{motioneq2} F(\phi)\tilde{g}^{AB}\tilde{\nabla}_A\tilde{\nabla}_B \phi= \zeta
\tilde{\mathcal{R}} \phi
+\tilde{V}^{\prime}(\phi)+\frac{1}{2}F^{\prime}(\phi)(\tilde{\nabla}\phi)^{2}
\end{eqnarray}
where the source term of the Ricci tensor is
\begin{eqnarray}
\label{newenmomdef}
\tilde{\mathcal{T}}_{AB}&\equiv&\tilde{T}_{AB}+\frac{1}{2-D}\tilde{g}_{AB}\tilde{T}\nonumber\\
&=&\tilde{\nabla}_A\phi\tilde{\nabla}_B\phi
F(\phi)-\zeta\tilde{\nabla}_A\tilde{\nabla}_B\phi^2-\frac{1}{2-D}\Big(2\tilde{V}(\phi)-
\zeta\tilde{\Box}\phi^2\Big)\tilde{g}_{AB}
\end{eqnarray}
and the energy-momentum tensor of the scalar field $\phi$ is
\begin{eqnarray}
\label{confenmom}
\tilde{T}_{AB}&=&\tilde{\nabla}_A\phi\tilde{\nabla}_BF(\phi)-\tilde{g}_{AB}\Big(\frac{1}{2}\tilde{g}^{CD}\tilde{\nabla}_C\phi\tilde{\nabla}_D\phi
F(\phi)+\tilde{V}(\phi)\Big)\nonumber\\
&+&\zeta\Big(\tilde{g}_{AB}\tilde{\Box}-\tilde{\nabla}_A\tilde{\nabla}_B\Big)\phi^2~~~.
\end{eqnarray}

It is obvious that the coupling of the scalar field to the curvature scalar induces the term with the zeta coupling in (\ref{confenmom}).

\subsection{Partial Vanishing of Energy-Momentum Tensor}
\label{subself1}

In this and the following subsection, we deal with (\ref{newenmomdef}), the source term of the Ricci tensor, in four dimensions 
and in extra dimensions to find the form of the scalar field $\phi$, its self-interaction potential and the curvature
scalar in extra dimensions. We analyze the partially 
gravitating scalar fields which gravitate only in extra dimensions. We
studied the non-gravitating scalar fields that have been already
analyzed in \cite{ayon-beato} and the
partially gravitating scalar fields in \cite{demir} to
help us for this work.

The first thing we do here is to put conditions on $\tilde{\mathcal{T}}_{AB}$, so it vanishes partially, \emph{i.e.}
$\tilde{\mathcal{T}}_{\mu\nu}=0, \tilde{\mathcal{T}}_{\mu
\bar{j}}=\tilde{\mathcal{T}}_{\bar{i}\nu}=0$ and
$\tilde{\mathcal{T}}_{\bar{i}\bar{j}}\neq 0$ and the corresponding
metric tensor structure should be $g_{\mu \nu}=\eta_{\mu \nu}, g_{\mu
\bar{j}}=g_{\bar{i}\nu}=0$ and
$g_{\bar{i}\bar{j}}=g_{\bar{i}\bar{j}}(\vec{y})$ as discussed in
\cite{demir}.

We write the source term of the Ricci tensor in four dimensions
\begin{eqnarray}
\label{newenmom4d}
\tilde{\mathcal{T}}_{\mu\nu}(\phi)=\tilde{\partial}_\mu\phi\tilde{\partial}_\nu\phi
F(\phi)-\zeta\tilde{\partial}_\mu\tilde{\partial}_\nu\phi^2
-\frac{1}{2-D}\Big(2\tilde{\mathcal{V}}(\phi)-\zeta
\tilde{g}^{\alpha\beta}\tilde{\partial}_\alpha\tilde{\partial}_\beta\phi^2\Big)\eta_{\mu\nu}
\end{eqnarray}
where the following expression for the self-interaction potential of the scalar field, $\phi$, have been introduced
\begin{eqnarray}
\label{newpot}
\tilde{\mathcal{V}}(\phi)\equiv\tilde{V}(\phi)-\frac{1}{2}\zeta
\tilde{g}^{\overline{i}\overline{j}}\tilde{\nabla}_{\overline{i}}
\tilde{\nabla}_{\overline{j}}\phi^2
\end{eqnarray}
to make all the terms look in four dimensions. From now on, we remind that all the terms which include the potential, 
$\tilde{\mathcal{V}}(\phi)$, depend on all of the coordinates.

We require $\tilde{{\mathcal{T}}}_{\mu \nu}=0$ for all
$\mu,\nu=0,1,2,3$ to generate a four-dimensional Minkowski
manifold with strictly flat coordinates, \emph{i.e.}  the flow of the energy-momentum tensor 
in the direction of the four dimensions should vanish.
The conditions on
$\tilde{{\mathcal{T}}}_{\mu \nu}$ are
\begin{eqnarray}
\label{newenmom1}
\tilde{\mathcal{T}}^{\mu\neq\nu}_{\mu\nu}=0=\tilde{\partial}_\mu\phi\tilde{\partial}_\nu\phi
F(\phi)-\zeta\tilde{\partial}_\mu \tilde{\partial}_\nu\phi^2~~~,
\end{eqnarray}
\begin{eqnarray}
\label{newenmom2}
\tilde{\mathcal{T}}_{00}=0=(\tilde{\partial}_0\phi)^2
F(\phi)-\zeta\tilde{\partial}^2_0\phi^2-\frac{1}{2-D}\Big(2\tilde{\mathcal{V}}(\phi)
-\zeta\eta^{\mu\nu}\tilde{\partial}_\mu\tilde{\partial}_\nu\phi^2\Big)\eta_{00}~,~~~\eta_{00}=-1
\end{eqnarray}
and
\begin{eqnarray}
\label{newenmom3}
\tilde{\mathcal{T}}_{ii}=0=(\tilde{\partial}_i\phi)^2
F(\phi)-\zeta\tilde{\partial}^2_i\phi^2-\frac{1}{2-D}\Big(2\tilde{\mathcal{V}}(\phi)
-\zeta\eta^{\mu\nu}\tilde{\partial}_\mu\tilde{\partial}_\nu\phi^2\Big)\eta_{ii}~,~~~\eta_{ii}=+1~~~.
\end{eqnarray}
We want to find the nontrivial configurations of the scalar field that nullify all components of
$\tilde{\mathcal{T}}_{\mu\nu}$ and the specific form of the
self-interaction potential which corresponds to these configurations of the scalar field.

We consider the ansatz for the scalar field
\begin{eqnarray}
\label{scalar} \phi(z) \equiv
\sigma(z)^{\alpha(z)}
\end{eqnarray}
where $\sigma(z)$ and $\alpha(z)$ also depend on all coordinates $z$. We make this ansatz to find the most general form of the scalar field
that satisfies the conditions (\ref{newenmom1}, \ref{newenmom2} and \ref{newenmom3}), so we have introduced the scalar field in terms of two different functions which depend on the coordinates $z$.

Firstly, we find the form of $\sigma$ by summing (\ref{newenmom2}) and (\ref{newenmom3})
\begin{eqnarray}
\label{sumenmom}
0&=&\tilde{\mathcal{T}}_{00}+ \tilde{\mathcal{T}}_{ii}\nonumber\\
&=&F(\phi) \sigma^{2\alpha}\Bigg(\alpha^2 u^{\prime 2} +2\alpha \alpha^\prime u^\prime 
+u^2 \alpha^{\prime 2} \Bigg) \Bigg((\partial_0 \sigma )^2 +(\partial_i \sigma)^2 \Bigg)\nonumber \\
&-& \zeta \sigma^{2 \alpha} \Bigg( 4u^2 \alpha^{\prime 2} + 8 u u^\prime \alpha \alpha^\prime + 4 \alpha^2 u^{ \prime 2} + 2 u \alpha^{\prime \prime}+ 4u^\prime \alpha^\prime - 2 \alpha u^{\prime 2} \Bigg) \Bigg((\partial_0 \sigma )^2 +(\partial_i \sigma)^2 \Bigg)\nonumber \\
&-&2 \zeta \sigma^{2 \alpha} \Bigg(\alpha u^\prime - u \alpha^\prime\Bigg) \Bigg(\partial_0^2 \sigma +\partial_i^2 \sigma \Bigg)
\end{eqnarray}
where we have introduced
\begin{eqnarray}
\label{sigmadef}
u\equiv\ln\sigma~~,~~~~u^\prime\equiv\frac{\partial\ln\sigma}{\partial\sigma}=\sigma^{-1}~~~.
\end{eqnarray}

From the last line of (\ref{sumenmom}) it is obvious that $\sigma(z)$ is a second order polinomial function
\begin{eqnarray}
\label{sigma} \sigma(z)= \frac{1}{2} a \eta_{\mu \nu} x^{\mu}
x^{\nu} + \eta_{\mu \nu} x^{\mu} p^{\nu} + b(\vec{y})
\end{eqnarray}
where $a$, $b$ and $p_{\mu}$ are integration constants. Here, $b(\vec{y})$ is, in general,
a function of extra coordinates.

Secondly, in order to find the explicit form of $\alpha(z)$ we analyze the first condition on $\tilde{\mathcal{T}}_{\mu \nu}$
\begin{eqnarray}
\label{enmomresult1}
0=\tilde{\mathcal{T}}^{\mu\neq\nu}_{\mu\nu}&=&\tilde{\partial}_\mu\sigma^\alpha\tilde{\partial}_\nu\sigma^\alpha
F(\sigma^\alpha)
-\zeta\tilde{\partial}_\mu\tilde{\partial}_\nu\sigma^{2\alpha}~~~.
\end{eqnarray}
We substitute the derivations 
\begin{eqnarray}
\label{scalardif4}
\tilde{\partial}_\mu \sigma^\alpha \tilde{\partial}_\nu \sigma^\alpha=\sigma^{2 \alpha} \Bigg(\alpha^2 u^{\prime 2}+2\alpha \alpha^{\prime}u u^\prime
+u^2 \alpha^{\prime 2} \Bigg) \tilde{\partial}_\mu \sigma \tilde{\partial}_\nu \sigma
\end{eqnarray}
and
\begin{eqnarray}
\label{scalardif5}
\tilde{\partial}_\mu \tilde{\partial}_\nu\sigma^{2\alpha}&=&2\sigma^{2\alpha}\Bigg(2u^2\alpha^{\prime 2}+4uu^\prime\alpha\alpha^\prime+2u^\prime\alpha^\prime
+2u^{\prime 2}\alpha^2+u\alpha^{\prime\prime}-u^{\prime 2}\alpha\Bigg)\partial_\mu\sigma\partial_\nu\sigma\nonumber\\
&+&2\sigma^{2\alpha}\Bigg(u^\prime\alpha+\alpha^\prime
u\Bigg)\tilde{\partial}_\mu \tilde{\partial}_\nu \sigma
\end{eqnarray}
into (\ref{enmomresult1}). Then, $\tilde{\mathcal{T}}_{\mu \nu}^{\mu \neq \nu}$ takes the form
\begin{eqnarray}
\label{enmomresult2}
0&=&\tilde{\mathcal{T}}^{\mu\neq\nu}_{\mu\nu}\nonumber\\
&=&\sigma^{2\alpha}\Big\{\Big(F(\sigma^\alpha)-4\zeta\Big)(u\alpha^\prime+u^\prime\alpha)^2\nonumber\\
&-&2\zeta(2u^\prime\alpha^\prime-u^{\prime
2}\alpha+u\alpha^{\prime\prime}+u^{\prime\prime}\alpha-u^{\prime\prime}\alpha)\Big\}
\tilde{\partial}_\mu\sigma\tilde{\partial}_\nu\sigma\nonumber\\
&=&\sigma^{2\alpha}\Big\{\Big(F(\sigma^\alpha)-4\zeta\Big)t^2-2\zeta
t^\prime+2\zeta(u^{\prime 2}\alpha+u^{\prime\prime}\alpha)\Big\}
\tilde{\partial}_\mu\sigma\tilde{\partial}_\nu\sigma\nonumber\\
&=&\sigma^{2\alpha}\Big\{\Big(F(\sigma^\alpha)-4\zeta\Big)t^2-2\zeta
t^\prime\Big\}
\tilde{\partial}_\mu\sigma\tilde{\partial}_\nu\sigma
\end{eqnarray}
where $t\equiv u\alpha^\prime+u^\prime\alpha$ and
$t^\prime\equiv\tilde{\partial} t /
\tilde{\partial}\sigma=u\alpha^{\prime\prime}
+2u^\prime\alpha^\prime+u^{\prime\prime}\alpha$. All primes refer to differentiations with respect to $\sigma$ and we have used $(u^{\prime 2}\alpha+u^{\prime\prime}\alpha)=0$ in the last line.

It is obvious from (\ref{enmomresult2}) that $\Big(F(\sigma^\alpha)-4\zeta\Big)t^2-2\zeta t^\prime=0$. We integrate this expression side by side two times.
Firstly, we integrate for $\sigma$ at one side and for t on the other side
\begin{eqnarray}
d\sigma=\frac{2\zeta}{\left(F(\phi)-4\zeta\right)t^2}dt~~,~~~\sigma=\frac{-2\zeta}{\left(\alpha^\prime\ln\sigma+\sigma^{-1}\alpha\right)\left(F(\phi)-4\zeta\right)}
\end{eqnarray}
Then, we integrate $\alpha$ terms at one side and $\sigma$ terms at the other side both for $\sigma$ and obtain $\alpha$
\begin{eqnarray}
\label{alpha}
\alpha = \frac{1}{\ln \sigma}\ \int^{\sigma}
\left(\frac{- 2 \zeta}{F\left(\widetilde{\sigma}^{\alpha}\right) -
4 \zeta} \right) \frac{d\widetilde{\sigma}}{\widetilde{\sigma}}
\end{eqnarray}
which is a function of $\sigma(z)$.

In the presence of $F\left(\phi\right)\neq 1$, the solution for
the field profile reads as in (\ref{scalar}) with (\ref{sigma})
where $a$, $b$ and $p_{\mu}$ are constants which are, in general,
functions of the extra coordinates. 

So, according to our ansatz (\ref{scalar}), the form of the scalar form is
\begin{eqnarray}
\label{specialphi}
\phi(z)=\Bigg(\frac{1}{2} a \eta_{\mu \nu} x^{\mu}
x^{\nu} + \eta_{\mu \nu} x^{\mu} p^{\nu} + b\Bigg)^{\frac{1}{\ln \sigma}\ \int^{\sigma}
\left(\frac{- 2 \zeta}{F\left(\widetilde{\sigma}^{\alpha}\right) -
4 \zeta} \right) \frac{d\widetilde{\sigma}}{\widetilde{\sigma}}}~~~.
\end{eqnarray}

The effects of
$F\left(\phi\right)\neq 1$ are collected in $\alpha(\sigma)$,
which reads as (\ref{alpha}) whose right-hand side is an indefinite integral over
$\widetilde{\sigma}$. One notices that for $F\left(\phi\right)
=1$, $\alpha = \frac{-2\zeta}{1 - 4 \zeta}$ follows automatically
\cite{ayon-beato}. One notices that this integral relation generalizes that of \cite{ayon-beato, demir}.

We integrate the right hand side of (\ref{alpha}) by substituting the function $F(\phi)$, (\ref{funcF}), 
to see the relations between the parameters better
\begin{eqnarray}
\label{alphaf}
\alpha(\sigma) = \frac{1}{\ln \sigma}\ \int^{\sigma}
\frac{- 2 \zeta}{ \frac{1}{1-\zeta
M_\star^{2-D}\widetilde{\sigma}^{2\alpha}}\left\{-12\frac{D-1}{D-2}\zeta^2M_\star^{2-D}\widetilde{\sigma}^{2\alpha}+\Big(1-\zeta
M_\star^{2-D}\widetilde{\sigma}^{2\alpha}\Big)^2\right\} -
4 \zeta}  \frac{d\widetilde{\sigma}}{\widetilde{\sigma}}
\end{eqnarray}
Then, this yields
\begin{eqnarray}
\frac{-12\frac{D-1}{D-2}\zeta^2 M_\star^{2-D}e^{2z}+\Big(1-\zeta
M_\star^{2-D}e^{2z}\Big)^2-
4 \zeta\left(1-\zeta
M_\star^{2-D}e^{2z}\right)}{-2 \zeta \left(1-\zeta
M_\star^{2-D}e^{2z}\right)}dz =\frac{1}{\sigma}d\sigma
\end{eqnarray}
where
\begin{eqnarray}
z\equiv \alpha(\sigma)\ln \sigma~~~.
\end{eqnarray}
Hence, the integration side by side gives the following relation
\begin{eqnarray}
\label{intresult}
\alpha(\sigma)=\frac{2\zeta}{4\zeta-1}\left[ 1-\frac{B\sigma^{2\alpha(\sigma)}}{4\zeta\ln\sigma}-\frac{A\ln(-1+B\sigma^{2\alpha(\sigma)})}{4B\zeta\ln\sigma}\right]
\end{eqnarray}
where, we have introduced
\begin{eqnarray}
A&\equiv&-12\frac{D-1}{D-2}\zeta^2M^{2-D}_\star\nonumber\\
B&\equiv&\zeta M^{2-D}_\star\nonumber~~~.
\end{eqnarray}
So, the parameters in the scalar field (\ref{specialphi}) must satisfy the condition (\ref{intresult}).

Hereby, we find the self-interaction potential of the scalar field $\phi$, (\ref{scalar}), 
by analyzing the second condition on $\tilde{\mathcal{T}}_{\mu \nu}$
\begin{eqnarray}
\label{t001}
0&=&\tilde{\mathcal{T}}_{00}\nonumber\\
&=&(\tilde{\partial}_0\sigma^\alpha)^2
F(\sigma^\alpha)-\zeta\tilde{\partial}^2_0\sigma^{2\alpha}-\frac{1}{2-D}\Big(2\tilde{\mathcal{V}}
(\sigma^\alpha)-\zeta\eta^{\mu\nu}\partial_\mu\partial_\nu\sigma^{2\alpha}\Big)\eta_{00}~,~~~\eta_{00}=-1\nonumber\\
&=&\frac{2}{2-D}\tilde{\mathcal{V}}(\sigma^\alpha)+2\zeta\sigma^{2\alpha}\Big(u\alpha^\prime+u^\prime\alpha\Big)\Big(\frac{4a}{D-2}+a\Big)\nonumber\\
&+&\frac{2\zeta}{D-2}\sigma^{2\alpha}\Big(2u^2\alpha^{\prime
2}+4uu^\prime\alpha\alpha^\prime\nonumber\\
&+&2u^{\prime
2}\alpha^2+u\alpha^{\prime\prime}
+2u^\prime\alpha^\prime-u^{\prime
2}\alpha\Big)\Big(2a\sigma+p^2-2ab\Big)~~~.
\end{eqnarray}
Eq. (\ref{t001}) yields the self-interaction potential as
\begin{eqnarray}
\label{pot}
\tilde{\mathcal{V}}(\sigma^\alpha)&=&2\sigma^{2\alpha-1}\zeta a\frac{1}{u^{\prime 2}}\nonumber\\
&\times&\Bigg\{\Big(u^\prime
u\alpha^\prime+u^{\prime
2}\alpha\Big)\frac{D+2}{2}+2u^2\alpha^{\prime 2}
+4uu^\prime\alpha\alpha^\prime+2u^{\prime 2}\alpha^2+u\alpha^{\prime\prime}+2u^\prime\alpha^\prime-u^{\prime 2}\alpha\Bigg\}\nonumber\\
&+&\sigma^{2\alpha}\zeta\Big(p^2-2ab\Big)\Bigg\{2u^2\alpha^{\prime
2}+4uu^\prime\alpha\alpha^\prime+2u^{\prime
2}\alpha^2+u\alpha^{\prime\prime}
+2u^\prime\alpha^\prime-u^{\prime 2}\alpha\Bigg\}
\end{eqnarray}
We make an abbreviation in (\ref{alpha})
\begin{eqnarray}
\label{alphaX} \alpha(\sigma)\equiv\frac{1}{\ln\sigma}X(\sigma)
\end{eqnarray}
and rewrite the potential as the following
\begin{eqnarray}
\label{potX}
\tilde{\mathcal{V}}(\sigma^\alpha)&=&2\sigma^{2\alpha-1}a\zeta\Big[\sigma
X^\prime\frac{D+2}{2}+2X^{\prime 2}\sigma^2
+X^{\prime\prime}\sigma^2\Big]\nonumber\\
&+&\sigma^{2\alpha-2}\zeta\Big(p^2-2ab\Big)\Big(2X^{\prime
2}\sigma^2+X^{\prime\prime}\sigma^2\Big)~~~.
\end{eqnarray}
Consequently, the potential takes the form
\begin{eqnarray}
\label{potfuncF}
\tilde{\mathcal{V}}(\phi)&=&8\phi^{\frac{2\alpha-1}{\alpha}}\frac{a\zeta^2\Big(D+2\Big)}{\Big(F(\phi)-4\zeta\Big)^2}\Big(\zeta
-\zeta_{\textrm{crit}}(\phi)\Big)\nonumber\\
&+&2\phi^{\frac{2\alpha-2}{\alpha}}\frac{\zeta^2\Big(p^2-2ab\Big)}{\Big(F(\phi)-4\zeta\Big)^2}\Big(F(\phi)+\phi^{\frac{1}{\alpha}}
F^\prime(\phi)\Big)
\end{eqnarray}
where
\begin{eqnarray}
\label{zetacrit}
\zeta_{\textrm{crit}}(\phi)=\frac{1}{4}\frac{F(\phi)D-2\phi^{\frac{1}{\alpha}}
F^\prime(\phi)}{D+2}~~~,~~F^\prime(\phi)\equiv\frac{\partial F(\phi)}{\partial\sigma}
\end{eqnarray}
is the critical value of $\zeta$ for which the theory becomes conformal. The potential (\ref{potfuncF}), which is felt by the scalar
field $\phi$ that has the special form as we have shown in (\ref{specialphi}), must have this form in order to have a compactification effect in the theory. $\zeta_{\textrm{crit}}(\phi)$ takes values in the range of 
the limit values of (\ref{zetacrit}) for $d\rightarrow 0$ and $d\rightarrow \infty$. One realizes that for $F(\phi)=1$ 
these values of $\zeta_{\textrm{crit}}(\phi)$ reduces to the $\zeta_{\textrm{crit}}(\phi)$ values in \cite{demir, ayon-beato} which are 
$1/6$ for $d\rightarrow 0$ and $1/4$ for $d\rightarrow \infty$.

The potential has two different minima, according to the sign of  $(\zeta-\zeta_{\textrm{crit}}(\phi))$, which correspond to uncompactified and compactified spacetime structures. This fact will be explained in Sec. \ref{subself2}. 

\subsection{Compactification of Extra Dimensions}
\label{subself2}

It is useful here to remember (\ref{phi}). It shows the relation
between the scalar field $\phi$ and the function $f(\mathcal{R}$).
We will see all the equations in the following sections still in
terms of $\phi$ but it should be kept in mind that they are also in
terms of $f(\mathcal{R}$) or $f^\prime(\mathcal{R})$ because of
(\ref{phi}). We discuss the compactification mechanism via the
scalar-tensor theory (\ref{actionnonmin}). Hereby we state that the
$f(\mathcal{R})$ theory causes the compactification because the
scalar-tensor theory is derived from it.

The equations of motion in extra dimensions are
\begin{eqnarray}
\label{motioneq3}
\tilde{\mathcal{R}}_{\overline{i}\overline{j}}=\frac{\tilde{\mathcal{T}}_{\overline{i}\overline{j}}(\phi)}{M_\star^{D-2}-\zeta\phi^2}
\end{eqnarray}
\begin{eqnarray}
\label{motioneq4}
F(\phi)\tilde{g}^{\overline{i}\overline{j}}\tilde{\nabla}_{\overline{i}}\tilde{\nabla}_{\overline{j}}\phi
&=&\frac{1}{2}F^\prime(\phi)
\tilde{g}^{\overline{i}\overline{j}}\tilde{\partial}_{\overline{i}}\phi\tilde{\partial}_{\overline{j}}\phi
+\frac{1}{2}F^\prime(\phi)\eta^{\mu\nu}\tilde{\partial}_{\mu}\phi\tilde{\partial}_{\nu}\phi +\zeta \tilde{R}\phi+\tilde{V}^\prime (\phi)\nonumber\\
&-&F(\phi)\eta^{\mu\nu}\tilde{\partial}_\mu\tilde{\partial}_\nu\phi~~~.
\end{eqnarray}
Here the source term of the Ricci tensor is
\begin{eqnarray}
\label{tij1}
0\neq\tilde{\mathcal{T}}_{\overline{i}\overline{j}}=F(\phi)\tilde{\partial}_{\overline{i}}\phi\tilde{\partial}_{\overline{j}}\phi
-\zeta\tilde{\nabla}_{\overline{i}}\tilde{\nabla}_{\overline{j}}\phi^2-\frac{1}{2-D}\Big(2\tilde{\mathcal{V}}(\phi)-\zeta\eta^{\mu\nu}
\tilde{\partial}_\mu\tilde{\partial}_\nu\phi^2\Big)\tilde{g}_{\overline{i}\overline{j}}~~~.
\end{eqnarray}

We will use (\ref{tij1}) and the equations of motion (\ref{motioneq3}, \ref{motioneq4}) to find the Ricci scalar. We first want to write
the explicit form of the term in the paranthesis in (\ref{tij1}). So, we find the following double-derivative of $\phi^2$ by using the ansatz
(\ref{scalar}) as $\phi=\sigma^\alpha$
\begin{eqnarray}
\eta^{\mu \nu} \tilde{\partial}_{\mu} \tilde{\partial}_{\nu} \sigma^{2\alpha}&=& \Bigg(4\alpha^{\prime 2}\sigma^{2\alpha} \ln^{2} \sigma
+8\sigma^{2\alpha-1}\alpha \alpha^{\prime}\ln \sigma +4\alpha^{\prime}\sigma^{2\alpha-1}\nonumber \\
     &+& 4 \alpha^{2}\sigma^{2\alpha-2}-2\alpha \sigma^{2\alpha-2}+2\alpha^{\prime \prime}\sigma^{2\alpha}\ln \sigma \Bigg) 
\eta^{\mu \nu}\tilde{\partial}_{\mu}\sigma
\tilde{\partial}_{\nu}\sigma \nonumber \\
     &+& \Bigg(2\alpha^{\prime}\sigma^{2\alpha}\ln \sigma+2\alpha \sigma^{2\alpha-1} \Bigg) \eta^{\mu \nu} 
\tilde{\partial}_{\mu} \tilde{\partial}_{\nu} \sigma
\end{eqnarray}
using the definition (\ref{alphaX})
\begin{eqnarray}
\label{Xderivative}
\eta_{\mu\nu}\tilde{\partial}_\mu\tilde{\partial}_\nu\sigma^{2\alpha}&=&2\sigma^{2\alpha-2}\left(2\sigma^2 X^{\prime 2}+\sigma^2 X^{\prime\prime}\right)\eta_{\mu\nu}\tilde{\partial}_\mu\sigma\tilde{\partial}_\nu\sigma+2\sigma^{2\alpha}X^\prime\eta_{\mu\nu}\tilde{\partial}_\mu\tilde{\partial}_\nu\sigma\nonumber\\
&=&2\sigma^{2\alpha-2}\left(2\sigma^2X^{\prime 2}+\sigma^2X^{\prime\prime}\right)\left(2a\sigma+p^2-2ab\right)+2\sigma^{2\alpha}X^\prime 4a
\end{eqnarray}
where $\eta_{\mu\nu}\tilde{\partial}_\mu\sigma\tilde{\partial}_\nu\sigma=-\left(ax_0+p_0\right)^2+\sum_i\left(ax_i+p_i\right)^2=2a\sigma+p^2-2ab$ and $\eta_{\mu\nu}\tilde{\partial}_\mu\tilde{\partial}_\nu\sigma=4a$.
We find the explicit form of the term in the paranthesis in (\ref{tij1}) by using (\ref{potX}) and (\ref{Xderivative})
\begin{eqnarray}
\label{enmompar}
2\tilde{\mathcal{V}}(\phi)-\zeta\eta^{\mu\nu}\tilde{\partial}_\mu\tilde{\partial}_\nu\phi^{2}=\frac{-4a\zeta^2\left(D-2\right)}
{F\left(\phi\right)-4\zeta}\phi^{\frac{2\alpha-1}{\alpha}}~~~
\end{eqnarray}
So, we use the relation (\ref{newpot}) for the second term in (\ref{tij1}) and (\ref{enmompar}) for the term in 
the parathesis to find the trace of this source term of Ricci tensor in the following form
\begin{eqnarray}
\label{tracetensor2}
\tilde{\mathcal{T}}=F(\phi)\partial_{\overline{i}}\phi\partial^{\overline{i}}\phi+2\Big(\tilde{\mathcal{V}}(\phi)-\tilde{V}(\phi)\Big)
-\Big(D-4\Big)\frac{4a\zeta^2\phi^{\frac{2\alpha-1}{\alpha}}}{F(\phi)-4\zeta}~~~.
\end{eqnarray}

We want to write the first term of (\ref{tracetensor2}) explicitly. For this purpose, we use the following differential rule 
\begin{eqnarray}
\label{difdef}
g^{\overline{i}\overline{j}}\nabla_{\overline{i}}\nabla_{\overline{j}}\phi^2=2g^{\overline{i}\overline{j}}\nabla_{\overline{i}}\phi
\nabla_{\overline{j}}\phi+2g^{\overline{i}\overline{j}}\phi\nabla_{\overline{i}}\nabla_{\overline{j}}\phi
\end{eqnarray}
and write the left-hand side of it from (\ref{newpot}). For the second term on the right hand side of (\ref{difdef}), we use the second
equation of motion (\ref{motioneq4}) by replacing the last term of it with
\begin{eqnarray}
\label{scalardif}
\eta^{\mu\nu}\tilde{\partial}_\mu\tilde{\partial}_\nu\phi
F(\phi)=F(\phi)\Bigg[\frac{\tilde{\mathcal{V}}^\prime(\phi)}
{\zeta}+2a\zeta\left(D-2\right)\frac{\frac{2\alpha-1}{\alpha}\phi^{\frac{\alpha-1}{\alpha}}\left(F(\phi)-4\zeta\right)
-\phi^{\frac{2\alpha-1}{\alpha}}F^\prime(\phi)}{\left(F(\phi)-4\zeta\right)^2}\Bigg]~~~.
\end{eqnarray}
So, the differential rule (\ref{difdef}) turns out to
\begin{eqnarray}
\label{rule}
-\frac{2}{\zeta}\Big(\tilde{\mathcal{V}}(\phi)-\tilde{V}(\phi)\Big)&=&
\Bigg( 2+ \frac{F^{\prime}(\phi)\phi}{F(\phi)}\Bigg) g^{\overline{i}\overline{j}}
 ハ \partial_{\overline{i}}\phi
 ハ \partial_{\overline{j}} \phi\nonumber \\
 ハ &+& \frac{F^{\prime}(\phi)\phi}{F(\phi)} \eta^{\mu \nu}\partial_{\mu} \phi \partial_{\nu} \phi
 ハ + \frac{2\zeta \mathcal{R} \phi^{2}}{F(\phi)} + \frac{2\phi \tilde{V}^{\prime}(\phi)}{F(\phi)}
 ハ -\frac{2\tilde{\mathcal{V}}^{\prime}(\phi)}{\zeta}\nonumber \\
 ハ &-&4a\zeta (D-2)\frac{\frac{2\alpha-1}{\alpha}\phi^{\frac{\alpha-1}{\alpha}}(F(\phi)-4\zeta)
 ハ -\phi^{\frac{2\alpha-1}{\alpha }}F^{\prime}(\phi)}{(F(\phi)-4\zeta)^{2}}~~~.
\end{eqnarray}

We see that the first term on the right hand side of (\ref{rule}) contains what we need for the first term in (\ref{tracetensor2}).
Hence, plugging the term we need from (\ref{rule}) in (\ref{tracetensor2}), we obtain the trace of the source of the
Ricci tensor in extra dimensions
\begin{eqnarray}
\label{tracetensor3}
\tilde{\mathcal{T}}&=&\frac{F(\phi)}{2F(\phi)+\phi
F^\prime(\phi)}\Bigg\{-\frac{2F(\phi)\Big(\tilde{\mathcal{V}}(\phi)-\tilde{V}(\phi)\Big)}
{\zeta}\nonumber\\
&-&F^\prime(\phi)\phi\eta^{\mu\nu}\partial_\mu\phi\partial_\nu\phi-2\zeta\tilde{\mathcal{R}}\phi^2-2\phi\tilde{V}^\prime(\phi)+F(\phi)
\frac{2\tilde{\mathcal{V}}^\prime(\phi)}{\zeta}\nonumber\\
&+&4a\zeta\Big(D-2\Big)\frac{F(\phi)}{\Big(F(\phi)-4\zeta\Big)^2}\Bigg[\frac{2\alpha-1}{\alpha}\phi^{\frac{\alpha-1}{\alpha}}\Big(F(\phi)
-4\zeta\Big)-F^\prime(\phi)\phi^{\frac{2\alpha-1}{\alpha}}\Bigg]\Bigg\}\nonumber\\
&+&2\Big(\tilde{\mathcal{V}}(\phi)-\tilde{V}(\phi)\Big)-\Big(D-4\Big)\frac{4a\zeta^2\phi^{\frac{2\alpha-1}{\alpha}}}{F(\phi)-4\zeta}~~~.
\end{eqnarray}

Finally, taking the trace of the first equation of motion (\ref{motioneq3}) 
\begin{eqnarray}
\label{trace2}
\tilde{\mathcal{T}}=\frac{\tilde{\mathcal{R}}}{M_\star^{D-2}-\zeta\phi^2}
\end{eqnarray}
and equaling the right hand sides of (\ref{tracetensor3}) and (\ref{trace2}), we find the Ricci scalar in the
extra dimensions
\begin{eqnarray}
\label{ricciscalar}
\tilde{\mathcal{R}}&=&\frac{1}{M_{\star}^{D-2}-\zeta
\phi^2+\frac{2 \zeta
\phi^2}{2+\phi\frac{F^{\prime}(\phi)}{F(\phi)}}} \Bigg\{
\frac{F(\phi)}{2F(\phi)+\phi F^{\prime}(\phi)} \Bigg[
-\frac{2F(\phi)}{\zeta}(\tilde{\mathcal{V}}(\phi)-\tilde{V}(\phi))\nonumber ハ\\
 ハ ハ &-& F^{\prime}(\phi)\eta^{\mu \nu} \phi
{\tilde{\partial}}_{\mu}\phi {\tilde{\partial}}_{\nu}\phi-2\phi
\tilde{\mathcal{V}}^{\prime}(\phi)+\frac{2F(\phi)}{\zeta}
\tilde{\mathcal{V}}^{\prime}(\phi)\nonumber \\
 ハ ハ &+& 4a\zeta (D-2)F(\phi)
\frac{\frac{2\alpha-1}{\alpha} \phi^{\frac{\alpha-1}{\alpha}}
(F(\phi)-4
\zeta)-\phi^{\frac{2\alpha-1}{\alpha}}F^{\prime}(\phi)}{(F(\phi)-4\zeta)^{2}}
\Bigg]\nonumber \\
 ハ ハ &+& 2(\tilde{\mathcal{V}}(\phi)-\tilde{V}(\phi))-\frac{4(D-4)a
\zeta^{2} \phi^{\frac{2\alpha-1}{\alpha}}}{F(\phi)-4 \zeta}
\Bigg\}
\end{eqnarray}
which is the expression that gives information about how extra
space is curved. In other words, it characterizes the curvature of the
extra space.

Here, we look into the properties of the potential
(\ref{potfuncF}) in detail. It is not the true self-interaction
potential of the scalar field but the potential that is felt by
the generic scalar field in four dimensions. The true self
interaction potential of $\phi$ is $\tilde{V}(\phi)$, \emph{i.e.} $(4+d)$-dimensional potential. The special form of
the potential, $\tilde{\mathcal{V}}(\phi)$, which satisfies the equations of motion,
(\ref{motioneq1}) and (\ref{motioneq2}), takes role in compactification process and keeps the four-dimensional spacetime flat. 
The warped compactified spacetime is energetically chosen structure of
spacetime instead of $M^{4+d}$ for specific values of $\phi$ that makes the potential
the potential $\tilde{\mathcal{V}}(\phi)$ minimum.

\begin{figure}[H]
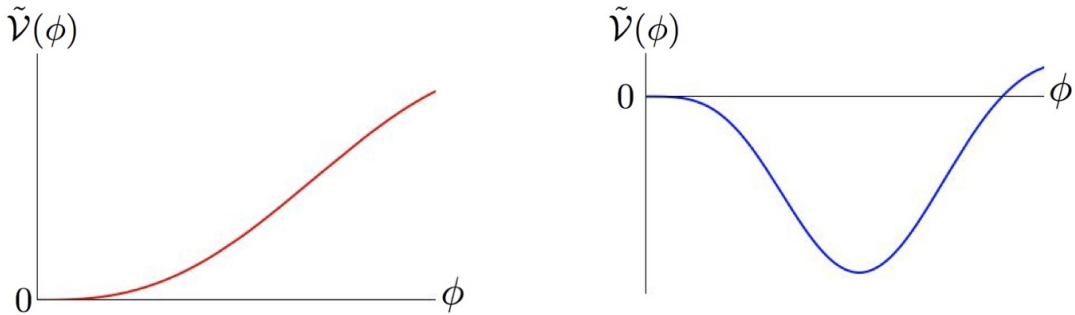

\centering
\mbox{\subfigure{\includegraphics[width=3in]{uncomp.jpg}}\quad
\subfigure{\includegraphics[width=3in]{comp.jpg} }}
\caption{Left: The minimum of the potential for $\zeta>\zeta_\textrm{crit}$ which corresponds to $M^{4+d}$. Right: The minimum of the potential for $\zeta<\zeta_\textrm{crit}$ which corresponds to the warped compactified spacetime.} \label{graphs}
\end{figure}

The potential (\ref{potfuncF}) have two minima at $\phi=0$ with $\zeta > \zeta_{\textrm{crit}}$ and $\phi \neq 0$
with $\zeta < \zeta_{\textrm{crit}}$. For both cases, $p^{2}-2ab>0$ and $a>0$ are taken. It is obvious that $\zeta$ values determines the
structure of the entire spacetime, \emph{i.e.} the structure
spontaneously changes from the uncompactified spacetime structure
$M^{4+d}$ to the warped compactified spacetime structure.

\section{Summary and Conclusion}
\label{conc}

We have shown a self-compactification mechanism via higher curvature gravity, $f(\mathcal{R})$. First, by a conformal transformation,
we have mapped $f(\mathcal{R})$ theory into Einstein gravity plus a scalar field theory with a minimal coupling and then by another conformal transformation we
have mapped the resulting theory to Einstein gravity plus a scalar field theory with 
a nonminimal coupling. We have shown that there are non-vanishing scalar field configurations that
satisfy the conditions on the partially vanishing source term, $\tilde{\mathcal{T}}_{AB}$, of the Ricci tensor. 

Compactification mechanisms were studied in the nonminimal scalar tensor theories. In these works, the Lagrangian of the theory is written by hand, and there is no any function, which depends on the scalar field which is the source of gravity, in front of the kinetic term. In our work, it is interesting and different that we didn't put our nonminimal scalar tensor action by hand, such that we derived it from a pure higher curvature gravity $f(\mathcal{R})$ via conformal transformations. We showed that, besides the nonlinear term which shows the coupling between the curvature scalar and the scalar field, our nonminimal scalar tensor theory includes also a nonminimal kinetic term. Hereby, we say that, if the thing which causes compactification is a higher curvature gravity, then the nonminimal scalar tensor theory which is obtained from the higher curvature gravity must have a non-minimal kinetic term with $F(\phi)$.

The scalar field floats in the bulk without coupling
to any field, however when the scalar field has a specific
configuration, it couples to gravity and this coupling term in the action plays a role for the compactification.
So, the scalar field gravitates only in a subset of dimensions, \emph{i.e.} in extra dimensions. This means that
extra space has a non-vanishing curvature scalar which we have already shown its form explicitly in (\ref{ricciscalar}). Curved space of extra dimensions may possess compact form \cite{cremmer, randjbar, omero, horvath} or not \cite{nicolai}.

The special form of the potential, $\tilde{\mathcal{V}}(\phi)$, which satisfies the equations of motion,
(\ref{motioneq1}) and (\ref{motioneq2}), determines the structure of the entire spacetime, \emph{i.e.} it plays role 
for the compactification of extra dimensions and keeps the four-dimensional spacetime flat. The warped compactified spacetime is the energetically 
chosen structure of the spacetime instead of $M^{4+d}$ for specific configurations of the scalar field $\phi$ and
the corresponding potential $\tilde{\mathcal{V}}(\phi)$. The solutions of the equations of motion, (\ref{motioneq3}, \ref{motioneq4}), give information about the topology and the shape of the extra space, but it is not easy to have an analytic solution since the equations of motion depend on functions 
of extra dimensions $b(\vec{y})$ and this function depends on $\tilde{g}_{\overline{i}\overline{j}}$ as already mentioned in \cite{demir}.

In our compactification mechanism all the results are in terms of the scalar field $\phi$, however it is important to keep in mind that according to the relation (\ref{scalar1}), all the results can be rewritten in terms of $f(\mathcal{R})$. So the whole mechanism is described only in terms of $f(\mathcal{R})$. Additionally, differently from other works, we realized that the scalar field which is the source of gravity is forced to have discrete spectrum via the equation (\ref{intresult}), and one obtains different spectrums for each different values of the parameters $D$, $\zeta$ and $M_\star$.

Ultimately, under all these illuminations, we have shown that $f(\mathcal{R})$ theory in $(4+d)$ dimensions can self-compactify extra dimensions while the four-dimensional spacetime remains flat. Manifestly, the result of the paper
is that the theory with non-canonical kinetic term (\ref{actionnonmin}) also accommodates
self-compactification. The relation to $f(\mathcal{R})$ gravity is new, however a product
compactification corresponds to a warped
compactification.

\section{Acknowledgements}

We thank to Prof.~Dr.~D.A.~Demir and Prof.~Dr.~V.V.~Nesterenko for illuminating discussions.
This work was partially supported by DFG GRK1102.


\end{document}